\documentclass[aps,prd,email,showpacs,twocolumn,preprintnumbers,showkeys,superscriptaddress,amsmath,amssymb,nofootinbib]{revtex4-2}

\usepackage{bbold}
\usepackage{color}
\usepackage{latexsym}
\usepackage{amsmath}
\usepackage{amssymb}
\usepackage[utf8]{inputenc}
\usepackage{amsmath}
\usepackage{amsfonts}
\usepackage{bm}
\usepackage{bbold}
\usepackage{float}
\usepackage{slashed} 
\usepackage{amssymb}
\usepackage{epstopdf}
\usepackage{epsfig}
\usepackage{eufrak}
\usepackage{euscript}
\usepackage{pstricks}
\usepackage{graphics}
\usepackage{graphicx}
\usepackage{picture}
\usepackage{appendix}
\usepackage{enumitem}
\usepackage{mathrsfs}

\newcommand{\be}{\begin{equation}}
\newcommand{\ee}{\end{equation}}
\newcommand{\ba}{\begin{eqnarray}}
\newcommand{\ea}{\end{eqnarray}}

\usepackage{hyperref}
\hypersetup{
    colorlinks=true,
    linkcolor=teal,
    filecolor=magenta,      
    urlcolor=teal,
    citecolor=magenta
}

\usepackage{orcidlink}

\begin{document}

\title{\Large{Re-assessing special aspects of Dirac fermions in presence of Lorentz-symmetry violation}}

\author{ João Paulo S. Melo \orcidlink{0000-0001-5213-5183}} \email{jpsm@cbpf.br}
\affiliation{Centro Brasileiro de Pesquisas F\'isicas, Rua Dr. Xavier Sigaud
150, Urca, CEP 22290-180, Rio de Janeiro, RJ, Brazil}
\author{José A.  Helay\"el-Neto \orcidlink{0000-0001-8310-518X}}\email{helayel@cbpf.br}
\affiliation{Centro Brasileiro de Pesquisas F\'isicas, Rua Dr. Xavier Sigaud
150, Urca, CEP 22290-180, Rio de Janeiro, RJ, Brazil}

\begin{abstract}
%\noindent

This paper focuses on additional inspections concerning the fermionic sector of the Standard Model Extension (SME). In this context, our main effort in this contribution is to investigate effects of Lorentz-symmetry violation (LSV) on the Klein Paradox, the \textit{Zitterbewegung} and its phenomenology in connection to Condensed Matter Physics, Atomic Physics, and Astrophysics. Finally, we discuss a particular realization of LSV in the Dirac equation, considering an asymmetry between space and time due to a scale factor present in the linear momentum of the fermion, but which does not touch its time derivative. We go further and extend the implications of this asymmetry in the situation the scale factor becomes space-time dependent to compute its influence on the kinematics of the Compton effect with the extended dispersion relation for the fermion that scatters the photon.

\end{abstract}

%\keywords{Lorentz-symmetry violation, Klein Paradox, \textit{Zitterbewegung}, Compton effect.}

\maketitle

\newpage

\pagestyle{myheadings}

\section{Introduction}

The Standard Model (SM) of Particle Physics represents a remarkable success in the human approach to describing Nature. Nevertheless, certain observations resist explanation within its framework. This compels us to surpass its boundaries for a more comprehensive understanding of Nature. Lorentz symmetry, a cornerstone of the SM, has endured for almost a century. Yet, even this deeply-rooted symmetry may reveal subtle deviations, needing to survive rigorous testing through high-precision measurements.
In the first instance,  one might think of Lorentz-symmetry violation (LSV) as a denial of the Theory of Special Relativity, whose underlying symmetry is described by the  proper orthochronous Lorentz group, SO(1,3). However, it is essential to clarify that this is not the case. Just as a uniform external magnetic field established in three-dimensional space breaks down rotational symmetry, characterized by the group $SO(3)$, some anisotropy in space-time corresponds to the violation of $SO(1,3)$ symmetry. The entire investigation within the scope of LSV is motivated by the quest to understand the possible origins of spacetime anisotropies present in the quantum vacuum. These origins could potentially indicate new physics beyond the SM. In fact, in scenarios involving LSV, researchers are also attempting to comprehend the very structure of the quantum vacuum, which exhibits electromagnetic properties such as birefringence or dichroism \cite{colladay1998l, kostelecky02c}. 

The discussion about a possible scenario with LSV was initially addressed to by Dirac in his quest of an aether, where the idea of a new QED with the existence of a background vector that would establish a preferred direction was put forward to understand whether this vector could reveal the nature of the ultraviolet divergences in quantum field theory \cite{dirac1951}. Consequences of that approach were explored in \cite{bondi1951}. After this period, the discussion returns in the $60$s with the considerations of a possible composite photon, all within a scenario of spontaneous Lorentz-symmetry (LS) breaking \cite{bjorken1963}. Shortly after, the possibility of the graviton being a Goldstone boson in the presence of a LSV background vector \cite{phillips1966, pavlopoulos1967} was discussed, and LSV was also contemplated in the framework of Particle Physics to analyze how it affects the velocity dependence of the muon lifetime \cite{redei1967}. These works were part of the pursuit of understanding possible situations in which LSV could have measurable effects.  In the $70$s, the context LSV comes into play  with the investigation of the beta function of a non-covariant Yang-Mills theory and shows that the fixed point occurs in a regime where LS is restored \cite{nielsen1978}. In the $80$s, in the context of Grand Unification, the discussion of the possibility  proton decay came to light  with a possibility that it occurs with slight deviations of LS \cite{ellis1980, zee1982}. The $80$s also witnessed the proof of the finiteness of $N = 4$ Yang-Mills theories at all orders of perturbative expansion. However, this proof was not manifestly covariant, which initiated a discussion about the extent to which LS could be taken as a fundamental and inviolable symmetry \cite{nielsen1982, chadha1983, nielsen1983}.

The Dirac's work and the subsequent articles marks the embryonic discussions that gained consistency with the Superstrings, and culminated in the current discussions on LSV. After the so-called first String Revolution, the seminal work by V. A. Kostelecký and S. Samuel came out and the authors showed, within the context of open strings, that there are tensor fields that may acquire nontrivial vacuum expectation values, condensing  in the vacuum and breaking LS \cite{kostelecky1989s}. This work serves as the starting point for the next phase of LSV investigation, which is more inspired by String Theory and effective models that attempt to discuss new physical scenarios beyond the SM. They also established a number of phenomenological constraints on LSV physics \cite{kostelecky1989p}.
Subsequently,  S. M. Carroll, G. B. Field, and R. Jackiw (CFJ) explored  limits on LSV  via a  $(1 + 3)$D version of the Chern-Simons topological  model in $(1 + 2)$D \cite{carroll1990}. In the CFJ model, there is a background vector field that establishes a spacetime anisotropy altering the Maxwell's equations, and they make use of astrophysical data to estimate the scale of its components. Some important considerations from the modified dispersion relation of the model arise, such as the effect of birefringence, and two aspects ought to be investigated, namely, the potential violation of causality and the stability property of the vacuum. We refer the reader to the works \cite{klinkhamer2001, helayel11}, where these issues are explored in detail.

The so-called Standard-Model Extension (SME) comes into play through works by D. Colladay, V. A. Kostelecký and others \cite{colladay1998l, colladay1997c}. The SME incorporates all the characteristics of the usual SM (i.e., the same local symmetries $SU(3) \times SU(2) \times U(1)$, being free from higher-order derivatives, and renormalizable) along with General Relativity but considers the possibility of violating Lorentz- and CPT-symmetry \cite{kostelecky04c}. In a sense, LS remains valid, as the theory behaves normally under rotations or boosts, considering that the so-called passive Lorentz transformations of the observer are performed in the laboratory frame. In the SME context, LSV only appears when the fields are rotated or altered with respect to the expected tensor values describing the vacuum state, known as active Lorentz transformations of particles.
Among the various contemporary approaches to investigate LSV, the SME is the most widely used theoretical framework to study phenomenology from multiple perspectives. The  SME incorporates the CFJ model and opens new avenues of exploration to demonstrate that there are situations where LSV effects can manifest at energy scales accessible in particle accelerators and through current or upcoming astrophysical and cosmological observations.

Soon afterwards, discussions related to LSV tests based on Modified Dispersion Relations (MDRs) for photons came to light  \cite{amelino1997, amelino1998, ellis1999probing, aloisio00}, which can contribute, using high-energy photons accessed in astrophysical measurements, in the analysis of possible deviations compared to standard dispersion relations indicating the presence of LSV or some effects of quantum gravity.
Additionally, in the context of MDRs, an important work by R. Gambini and J. Pullin came out, which did not rely on ST or the usual quantum gravity formalism, such as first-order formalism or semiclassical approach. Instead, it was based on the Loop Quantum Gravity (LQG) framework. They obtained modified Maxwell equations and demonstrated vacuum birefringence \cite{gambini1999}. Two other important works regarding MDRs, characterizing dispersion relations that could indicate the existence of LSV, are \cite{amelino2000w, alfaro2002l}.

In $1999$, N. Seiberg and E. Witten introduced the Non-Commutative Field Theories, {which exhibit LS breaking}, and it can be seen as a low-energy limit of ST \cite{seiberg1999}. This  approach to investigating LSV can be found in the literature and a very seminal work in this regard is \cite{caroll01}. Also S. Coleman and S. L. Glashow introduced, from first principles, the hypothesis that there could be LSV at extremely high energies, near the Planck scale $\sim 10^{19}$ GeV \cite{coleman1999}. Later, came another approach to LSV, known as Doubly-Special Relativity (DSR), which involves a theory of double relativity, in the sense that there are two invariants: the speed of light and an energy scale, equivalent to a minimum length associated with the Planck scale \cite{amelino2002, magueijo2003}. Then, there is a direct LSV in the algebra of the Poincaré group due to the presence of this minimum invariant length scale. This modification of the Poincaré group algebra is achieved through a mathematical procedure called Wigner-In\"on\"u contraction. Starting from an Anti-de Sitter group, a Wigner-In\"on\"u contraction is performed, resulting in the so-called $\kappa$-Poincaré algebra. This new algebra is a deformation of the usual Poincaré algebra, characterized by a length scale  that modifies the well-known LS \cite{Kowalski-Glikman04}.

With all this background provided,  it could be stated that LSV investigation implies that LS is an indispensable fact of nature at the scales we live but the question of to what extent can we uphold this symmetry is an open issue. We have good reasons to believe that it should break at extremely short distance scales, this is at ultra-high energies. Therefore, in low energies, we do not expect to observe a complete LSV, but rather its manifestations through very subtle effects, as they can be suppressed by some power of the Planck scale.
Experimental and observational searches for these subtle effects have yielded good limits on the parameters that govern LSV. The most recent results are published annually  in the so-called ``\textit{Data Tables for Lorentz and CPT Violation}''  \cite{russel2020}. Another interesting work  provides a general review of LSV phenomenology and gives indications on what data to search for to establish limits on LSV \cite{bolmont2020}. These searches generally focus on evaluating phenomenologies related to the motion of objects in an anisotropic background \cite{bluhm02, bluhm03}, atomic clocks in space \cite{Delva17, kostelecky18, Sanner19}, vacuum birefringence of light \cite{Kostelecky01b, Kostelecky08b, Kostelecky13b, Meuren17}, spin precession effects \cite{Bennett08, Allmendinger14, Kostelecky14s}, atomic energy level shifts and spectroscopy \cite{Kostelecky16p, Dreissen22}, neutrino physics \cite{Amelino-Camelia15, Amelino-Camelia16, Diaz16a, Diaz16b, Amelino-Camelia17, Huang18, Huang19, Huang22}, modified dispersion relations for high-energy photons and observations of gamma-ray bursts \cite{amelino1997, Kostelecky13b, Shao10, Zhang15, Xu16a, Xu16b, Liu18, Li20, Zhu21, Amelino-Camelia21}, among others \cite{Gomes16, Schreck17, Araujo19, Gomes22, Crivellin22, Fabritiis22a, Fabritiis22b, Petrov22, Qin23, Melo24}, establishing constraints on the parameters which characterize them.

The topic of LSV is a pursuit of new physics that goes beyond  the SM, with the hope of shedding light on various questions such as: indicating a path towards quantum gravity (ST, LQG, etc.); addressing the hierarchy problem in gauge theories; understanding the origin of the Higgs boson mass; explaining the asymmetry between matter and antimatter in the universe; resolving the neutron electric dipole moment problem; dark matter; investigating neutrino oscillations and their mass;  among others. 
 In this regard, a  recent work that  marks the intersection between LSV and Condensed Matter Physics (CMP), opening a very promising line of investigation in establishing a connection between the field and Dirac and Weyl semi-metals, offering the prospect of new laboratory experiments with materials that could provide new regions of validity and new limits on the parameters associated with LSV \cite{kosteleckyCM}. This new direction in the field is highly auspicious since it will create another channel of communication between the High Energy Theory and the CMP communities. 

Given all this background about the LSV, in Sect. \ref{sec1}, we propose to re-assess properties of the Dirac equation in the SME context, which leads us to better comprehend the LSV effects on the Klein Paradox and on the phenomenology of \textit{Zitterbewegung}, which have relations with Condensed Matter Physics, Atomic Physics, and Astrophysics. In Sect. \ref{sec2}, we discuss a novel form of LSV in the Dirac equation, disrupting the hierarchy between space and time due to a scale factor that multiplies the spatial part of the four-derivative. The intention is to extend the implications of this type of symmetry, especially when it exhibits a local character and its influence on the Compton effect kinematics.  Finally, in Sect. \ref{sec4} we make our concluding remarks. We adopt the Minkowski signature $(+, -, -, -)$ thoughout this text.

\section{Inspections in the fermionic sector of SME}
\label{sec1}

 The QED sector of the SME  consists in the following action in flat spacetime \cite{Kostelecky1}:
\begin{align}
S_1 =& \int d^4 x \bigg[  \bar{\psi}\left( i\hbar \Gamma^\mu D_{\mu} -M c\right)\psi -\dfrac{1}{4\mu_0} F^{\mu\nu}F_{\mu\nu} \nonumber \\
    & - \!  \dfrac{1}{4\mu_0} {( k_{F})}^{\mu\nu\lambda\rho}F
    _{\mu\nu}F_{\lambda\rho} \! - \! \dfrac{1}{2\mu_0}\varepsilon_{\mu\nu\kappa\lambda}{(k_{AF})}^{\mu}A^{\nu} F^{\kappa\lambda}   \bigg] , \label{actionnn}
\end{align} 
where the dimensionless constant background tensor $k_{\mu\nu\lambda\rho}$ is a CPT-even term that carries similar symmetry properties of curvature tensor in gravitation context, and the total term that contracts with it is commonly referred to as the aether term. The constant background vector $k^{\mu}$ is a CPT-odd term with dimensions of inverse of length, and the total term that accompanies it is the usual CFJ term. $D_{\mu} = \partial_\mu +ie A_\mu$ is the usual covariant derivative. We also have 
\begin{align}
\Gamma^\mu \!&= \! \gamma^\mu \!+ \!c^{\mu\nu}\gamma_\nu \!+\!d^{\mu\nu}\gamma_\nu\gamma_5 \!+\! e^\mu \!+\!  i f^\mu\gamma_5  \!+\! \dfrac{1}{4}g^{\lambda\nu\mu}\Sigma_{\lambda\nu} , \label{GGterm} \\ 
M          &= m + im_5 \gamma_5 + a_\mu \gamma^\mu + b_\mu \gamma^\mu\gamma_5  +\dfrac{1}{4} H_{\mu\nu}\Sigma^{\mu\nu} , \label{MMterm}
\end{align}
with $\gamma^\mu$ and $\gamma_5 = i\gamma^0\gamma^1\gamma^2\gamma^3$ being the usual Dirac matrices obeying the Clifford algebra $\left\{\gamma^\mu, \gamma^\nu \right\} = 2\eta^{\mu\nu}$, and  defining the $SO(1,3)$ generators in  the spinor representation, $\Sigma^{\mu\nu} =i[\gamma^\mu,\gamma^\nu]/4$. The terms in Eqs. \eqref{GGterm} and \eqref{MMterm} can be written as $\Gamma^\mu =\gamma^\mu + \delta \Gamma_{\textrm{\tiny LSV}}^\mu $  and  $M= m + \delta M_{\textrm{\tiny LSV}} $, with both $\delta\Gamma_{\textrm{\tiny LSV}}^\mu $ and $\delta M_{\textrm{\tiny LSV}} $  terms, which are, respectively, dimensionless and with dimensions of  mass, being  small LSV and CPT violating terms. Imposing that the Lagrangian ought to be Hermitian, implies that all of these LSV coefficients are real. The $a_\mu$, $b_\mu$, $e_\mu$,  and $g_{\mu\nu\lambda}$ are all CPT-violating, while the $f_\mu$ is CPT-even since it can be mapped to the $c_{\mu\nu}$ coefficients by a suitable spinor redefinition and it only occurs as bilinear combinations  $f_\mu f_\nu$ in physical
observables, see \cite{altschul06} for more details.

Regarding the Lorentz and CPT invariance of the action \eqref{actionnn}, one should keep in mind that these transformations are defined by the free field theory, $\mathcal{L}_0 = \bar{\psi}(i\hbar \gamma^\mu \partial_\mu -mc)\psi$. 
%and then applying it to the study of symmetry properties  of the action \eqref{actionnn}. 
Adopting that point of view, the difference between observer and particle transformations turns out to be important and then LSV terms  ${(k_{F})}^{\mu\nu\lambda\rho}$, ${(k_{AF})}^{\mu}$, $\delta\Gamma_{\textrm{\tiny LSV}}^\mu $ and $\delta M_{\textrm{\tiny LSV}} $  all transforms as scalar, four-vectors and tensors  under observers transformations and just as scalars under particles transformations. That distinction enables to define all LSV terms terms as constants background fields \cite{colladay1997c}. 

In what follows, we shall not consider the Maxwell, CFJ, aether  and $\delta\Gamma_{\textrm{\tiny LSV}}^\mu $ terms,  focusing just on the  mass sector, Eq. \eqref{MMterm}, considering $H_{\mu\nu} =0$. These simplifications are motivated by two factors:
i) We are neglecting contributions from more complex LSV backgrounds, such as those described by tensors.
ii) Considering potential future applications and implications of LSV in the low-relativistic limit, we recognize that the mass sector of the Dirac equation is dominating over the linear momentum in this regime.  
Therefore, we shall work with a modified Dirac equation that corresponds to the action below: 
\begin{equation} 
 S_2 \! = \! \int \! d^4 x \bar{\psi} \!  \left( i\hbar \gamma^{\mu} \partial_{\mu} \! - \! mc  \! - \! im_5 c\gamma_5 \!  - \! ca_{\mu}\gamma^{\mu}  \! - \!  cb_{\mu}\gamma^{\mu}\gamma_5 \right)\! \psi . \label{01}
\end{equation}

{
This is the right point to emphasize that, in some special situations, a field reparametrization can be performed in such a way that the Lorentz- and/or CPT-breakings can be removed; in other words, they can be suitably reabsorbed into a $\psi$-field redefinition. In Eq. \eqref{01}, that is the case of $a_\mu$, in general. As for $b_\mu$, it would also be so, but only in the absence of both the fermion mass, $m$, and the $m_5$-parameter. Field reshufflings, as given by $\psi (x) =e^{- ica\cdot x /\hbar}  \chi (x)$ and $\psi (x) = e^{-icb\cdot x \gamma_5/\hbar}  \chi (x)$, respectively, remove away the parameters in question}.

Taking the Fourier transform
\begin{align}
          \psi(x) = \int \dfrac{d^4 p}{(2\pi\hbar)^4} \psi_0(p)e^{-ip_{\mu}x^{\mu}/\hbar},
\end{align}
one can write the modified Dirac equation associated with Eq. \eqref{01} in the form 
\begin{align}
\left( p_{\mu} \gamma^{\mu} \!  - \! mc \!  - \! im_5c\gamma_5 \! - \! ca_{\mu}\gamma^{\mu} \! - \!  cb_{\mu}\gamma^{\mu}\gamma_5 \right)  \psi_0(p) =0\textrm{.} \label{02}
\end{align}

The modified fermion propagator in the momentum space associated with this dynamical equation is \cite{reis17, reis19}
\begin{widetext}
    \begin{align}
iS(p) =&\; i \Big\{  \Big(\slashed{p}- c\slashed{a} - c\slashed{b}\gamma_5 +mc-im_5c \gamma_5\Big)    \left[(p- c a)^2-(m^2+m_5^2+ b^2)c^2 \right] - 2 c \Big[ \left(\slashed{p}- c \slashed{a} \right)\gamma_5  + c \slashed{b} \Big] \big(p - c a \big)  \cdot b \nonumber \\
 &\;  -2 m_5 c^2\Big[\big(p_\mu-ca_\mu \big)b_\nu-\big(p_\nu- ca_\nu \big)b_\mu \Big]\Sigma^{\mu\nu} - 2 mc^2\; \varepsilon_{\alpha\beta \mu\nu} \Big[\big(p^\alpha-ca^\alpha \big)b^\beta-\big(p^\beta-c a^\beta \big)b^\alpha \Big]\Sigma^{\mu\nu}  \Big\} \Delta^{-1},
\end{align}  
\end{widetext}
with 
\begin{align}
\Delta =&\; \left[(p-ca)^2-(m^2c+m_5^2+b^2)c^2\right]^2  \nonumber \\
 &\; +4c^2b^2(p-ca)^2-4 \left[c b\cdot(p-ca)\right]^2. \label{deltaRD}
\end{align}
From this result, we can read off the modified dispersion relation from the poles of the propagator, this is,  taking $\Delta =0$ and solving it for energy in function of linear momenta. From Eq. \eqref{deltaRD}, without any
special choice, one can simply write the following implicit energy and momentum function:
\begin{align}
&  \left[(p- c a)^2-(m^2+m_5^2+b^2)c^2\right]^2  \nonumber \\
&  +4c^2b^2(p-a)^2-4 \left[c b\cdot(p-ca)\right]^2 =0 . \label{RDgeral}
\end{align}
From Eq. \eqref{RDgeral} turns out that solutions of the type $E(\bm p)$ are not analytically transparent  (more details of search for interesting special cases can be found in \cite{colladay1997c}). Even without an explicit solution, one can argue that the energy spectrum must be real, since the Hamiltonian
\begin{align} \label{Hamiltonian}
    H =&\; -i\hbar c \gamma^0 \bm \gamma \cdot \bm \nabla + c^2a_\mu \gamma^0\gamma^\mu  \nonumber \\ 
       &\; + mc^2 \gamma^0+im_5 c^2 \gamma^0\gamma_5 +c^2 b_\mu \gamma^0\gamma^\mu \gamma_5, 
\end{align}
which describes  modified Dirac equation associated with Eq. \eqref{01} through  $H \psi (x)= i \hbar \partial_t \psi (x)$, 
is Hermitian, see \cite{kostelecky99c}. 

As a consequence of Eq. \eqref{RDgeral}, one can also obtain the group velocity considering the  total differential of it,  
$
        d\Delta =  (\partial \Delta /\partial E)  dE + (\partial \Delta /\partial p_i)  d p_i = A dE + B dp_i =0 .
$
Then, the components of the group velocity is given by 
\begin{align}
 v_i  &= \dfrac{\partial E}{\partial p_i} = -\dfrac{B}{A} =\dfrac{c^2[d_1(p_i-ca_i)+d_2 cb_i]}{d_1(E-a_0 c^2)+d_2 b_0 c^2}, \label{gpvelo}
\end{align}
where $d_1=(p-ca)^2+(b^2-m^2-m_5^2 )c^2$ and $d_2=-2c(p-c a)\cdot b$. It reduces to the usual situation, $v_i =c^2 p_i/E$, if we turn off the LSV background. Let us notice that, even in the presence of LSV terms, the expressions for the group velocity admits the condition $|v_i| < c$, indicating in this case that the causal structure of the theory is maintained (see, e.g., \cite{helayel11, colladay1997c,AdaKli01, klinkhamer11, schreck12} for discussions on causality in scenarios with Lorentz violation).

The model admits both vector current, $j^\mu= \bar{\psi}\gamma^\mu \psi$ which is conserved, and chiral current, $j^\mu_5= \bar{\psi}\gamma^\mu \gamma_5\psi$ which is not conserved due to mass $m$. Then, one can couple these currents minimally with photon field, $A_\mu$, in order to read off the respective vertex structure. In momentum space, whether we calculate the Gordon decomposition, we can express these vertices in the form   $\bar{\psi}_0(p')  \gamma^\nu  \psi_0(p) {A_0}_\nu(q)  = \bar{\psi}_0(p')  \Theta^\nu  \psi_0(p) {A_0}_\nu(q)$ and $i\bar{\psi}_0(p')  \gamma^\nu \gamma_5 \psi_0(p) {A_0}_\nu(q)  = \bar{\psi}_0(p')  \tilde{\Theta}^\nu  \psi_0(p) {A_0}_\nu(q)$, with 
\begin{align} \label{eqDG1}
 \Theta^\nu =  \dfrac{1}{2mc} \Big[ \big(\ell^\nu-2ca^\nu\big)  - 2iq_\mu  \Sigma^{\mu\nu} 
- 4icb_\mu  \Sigma^{\mu\nu}\gamma_5  \Big] 
\end{align}
and
\begin{align}
\tilde{\Theta}^\nu= - \dfrac{1}{2m_5c} \big[ q^\nu  +cb^\nu \gamma_5  -2i\Big(\ell_\mu -2ca_\mu \Big) \Sigma^{\mu\nu}  \Big] , 
\end{align}
with $q_\mu = p'_\mu -p_\mu$ and $\ell_\mu =p'_\mu -p_\mu $ being the transfer and total momentum, respectively, and $A_{0\nu}$ the photon field in momentum space. 
At first glance, an interesting fact in these results is that in Eq. \eqref{eqDG1} we can see the vector current decomposed into the usual terms of the Gordon decomposition, and we can notice the emergence of an extra term accompanying the LSV term $b_\mu$, contracted with the dual form of the spin current, suggesting that the LSV induces the appearance of an electromagnetic form factor associated with a possible contribution to the electric dipole moment for the electron (EDM) \cite{Roberts10, Roussy23}.

Additionally,  the on-shell canonical energy-momentum tensor conservation law of the model is such that
\begin{align} \label{EMT1}
\partial_\mu T^{\mu\nu} = \partial_\mu \Big[ \dfrac{i\hbar}{2}\bar{\psi}\gamma^\mu(\partial^\nu \psi) -\dfrac{i\hbar}{2} (\partial^\nu \bar{\psi}) \gamma^\mu \psi   \Big] = 0 . 
\end{align}
In order to write the conservation law of a symmetric fermionic energy-momentum tensor, i.e.,  
\begin{align}
\theta^{\mu\nu}\! \! = \! \dfrac{i\hbar}{4} \! \big[ \bar{\psi}\gamma^\mu (\partial^\nu\psi)  \! + \! \bar{\psi}\gamma^\nu (\partial^\mu \psi)  \! - \! (\partial^\nu \bar{\psi})\gamma^\mu \psi \! - \! (\partial^\mu \bar{\psi})\gamma^\nu \psi  \big]
\end{align}
with $\partial_\mu \theta^{\mu\nu}=0$ for Lorentz invariant on-shell usual Dirac theory,  it is necessary to subtract the anti-symmetric part of the canonical conserved energy-momentum tensor in Eq. \eqref{EMT1}. Following the methodology  given by \cite{Freese22}, it is straightforward to show that in the present model this procedure lead us to write 
\begin{align} \label{EMT2}
   \theta^{\mu\nu}_{\textrm{\tiny LSV}} = \theta^{\mu\nu} + \dfrac{c}{2}\big[a^\mu,j^\nu \big]   + \dfrac{1}{2}\big[b^\mu,j^\nu_5 \big].
\end{align}
From Eq. \eqref{EMT2}, due to the presence of the CPT violating, we have that two important aspects regarding the symmetrization of $T^{\mu\nu}$ stands out: these terms makes the symmetrization impossible and $\theta^{\mu\nu}_{\textrm{\tiny LSV}}$ is no longer a conserved quantity.   

As argued by Kostelecký in \cite{kostelecky04grav}, in Minkowski spacetime  the LSV coefficients are constants and this fact simplifies physical consequences such as energy and momentum conservation, as we can see from Eq. \eqref{EMT1},  and plenty of physical arguments can be used to justify this assumption. As an example, certain explanations propose that higher energies are associated with coefficients exhibiting nontrivial spacetime dependence, which implies a natural preference for constant coefficients. In a broader context, if Lorentz breaking originates at the Planck scale and an inflationary period occurs in cosmology, it becomes plausible for present-day configurations to exhibit constant coefficients over the Hubble radius. Additionally, assuming sufficiently slow spacetime variation of the coefficients, the notion of constant LSV terms can be considered the primary approximation in a series expansion. It is important to note that all such arguments involve inherent physical choices. From a formal standpoint, any vector or tensor field with smooth integral curves is considered a valid candidate. Hence, opting for constant coefficients in Lorentz violation can be seen as a type of choice for the theory. 

This condition of constant LSV terms in a scenario of a Riemann-Cartan spacetime, implies integrability conditions  which are fulfilled globally only for special spacetimes, i.g., parallelizable manifolds, which have zero curvature and are comparatively rare in four or more dimensions with no interest for theories of gravity \cite{kostelecky04grav}. That said, it is reasonable to assume, at least in some region of the spacetime, a non-constant LSV background. Adopt this premise brings nontrivial consequences for the canonical energy-momentum tensor conservation law. This  is, 
\begin{align} 
\dfrac{\partial_\mu T^{\mu\nu}}{c} \! = \! j_\mu \partial^\nu a^\mu (x)  +  {j_5}_\mu \partial^\nu b^\mu (x)+ i\bar{\psi}\gamma_5 \psi \partial^\nu m_5(x).
\end{align}
This result indicates that the  LSV canonical energy-momentum tensor is just conserved upon the assumption that LSV terms are held constant, and only in this situation the translation invariance of the theory is preserved.

\subsection{Modified Dirac equation whenever $b_{\mu} =0$}

In this Section, our discussion will center around the scenario where $b_{\mu} =0$. While it may initially appear restrictive, concentrating in this particular case can open up the way for establishing constraints on the parameters, such as the $m_5$ parameter via Condensed Matter Physics data. With this purpose, we shall begin by presenting the comprehensive analytical solution to the modified Dirac equations. Subsequently, we shall go over to an examination of the Klein's paradox and devote some time to discuss the \textit{Zitterbewegung} phenomenology within this context.

Then, by setting  $b_{\mu} =0$, the  Eq. \eqref{01} assumes the form as given below:
\begin{align}
    \left(i\hbar \gamma^\mu \partial_\mu -mc - i m_5c \gamma_5 - ca_{\mu} \gamma^\mu \right) \psi(x) = 0 . \label{simplesDirac}
\end{align}
In momenta space 
\begin{align}
\left( p_{\mu} \gamma^{\mu}  -mc -im_5 c \gamma_5 -ca_{\mu}\gamma^{\mu}  \right)\psi(p) =0. \label{mmomspac}
\end{align}

Considering Eq. \eqref{RDgeral} for {$b_{\mu} =0$}, the dispersion relation can be expressed as 
\begin{align}
  E = a_0c^2 + c\sqrt{ \left(\bm{p} -c\bm{a} \right)^2 + m^2c^2  +m_5^2 } .
\end{align}
With the energy-momentum relation
\begin{equation}
       \begin{cases}
       \mathscr{E}  \equiv E - a_0 c^2   ,   \\ 
\bm{P} \equiv\bm{p} -c\bm{a} ,
\end{cases}\label{06}
\end{equation}
one can write
\begin{align}
\mathscr{E} =  c\sqrt{\bm{P}^2 +m^2 c^2+m_5^2c^2}  .
\end{align}

Now, whether one again consider the group velocity, one gets from \eqref{gpvelo}, when {$b_\mu = 0 $}, the result 
\begin{align}
 \bm{v} =  \dfrac{c^2 \left(\bm{p}-c\bm{a}\right)}{\left(E-a_0c^2\right)}= \dfrac{c^2\bm{P}}{\mathscr{E}} .
\end{align}

The modified fermion propagator in this case is \cite{reis17, reis19}
\begin{align}
iS\left( P \right) &= \dfrac{i\left(\slashed{p} -c\slashed{a} +mc -im_5 c\gamma_5  \right)}{\left(p^2 -2c p\cdot a +a^2 c^2 -m^2c^2 -m_5^2 c^2 \right)}  \nonumber \\
 &=\dfrac{i\left(P_\mu \gamma^\mu  +mc -im_5 c\gamma_5  \right)}{\left(P^2 -m^2c^2 -m_5^2 c^2+i\varepsilon \right)} , \label{propagator1}
\end{align} 
where we define an effective four-momentum given by
\begin{align}
P_\mu \equiv \left( \dfrac{\mathscr{E} }{c }, - \bm P\right).
\end{align}
This propagator is quite similar to usual fermion propagator, with exception of terms proportional to $m_5$.

The results for the Gordon decomposition and energy-momentum tensor, along with their respective interpretations, remain consistent with those presented in the previous section when we consider {$b_\mu = 0 $}. However, we will not utilize them in the subsequent analysis, and this is why we only mention them briefly.

\subsection{Working out positive- and negative-energy solutions}

With these remarks in mind, we are now prepared to delve into the study of solutions for the modified Dirac equation defined by equation \eqref{mmomspac}.
Starting off with the solutions for positive energies and spin-up ($+s$) in the laboratory frame  and working with the Dirac representation for the gamma matrices, the solutions take the following form
\begin{align}
\psi_{+}(P,+s) &= u(P,+s) e^{-iP_{\mu}x^{\mu}/\hbar}  \nonumber \\
  &=N_{+} \begin{pmatrix} 1 \\ 0 \\[0.1cm]
  \dfrac{c\left(P_3 -im_5c \right)}{\varepsilon +mc^2}  \\[0.3cm]  \dfrac{c\left(P_1+iP_2  \right)}{\varepsilon +mc^2} \end{pmatrix} e^{-iP_{\mu}x^{\mu}/\hbar}, \label{pspinup}
\end{align}
where
\begin{align}
 \varepsilon \equiv c\sqrt{\bm{P}^2 +m^2c^2 +m_5^2c^2} ,
\end{align}
and $N_+$ is a normalization constant to be determined.

In the case of positive energies and spin-down, $(-s)$,  we have 
\begin{align}
\psi_{+}(P,-s) &= u(P,-s) e^{-iP_{\mu}x^{\mu}/\hbar} \nonumber \\
&= N_{-} \begin{pmatrix} 0 \\ 1 \\[0.1cm] \dfrac{c\left(P_1-iP_2 \right)}{\varepsilon  +mc^2}  \\[0.3cm]   \dfrac{-c\left(P_3+im_5 c \right)}{\varepsilon  +mc^2} \end{pmatrix}  e^{-iP_{\mu}x^{\mu}/\hbar} ,
\end{align}
and once again $N_{-}$ is a normalization constant to be  determined.

The determination of both $N_+$ and $N_{-}$ comes from the exigence of $ \bar{\psi}_{+}(P,\pm s) \psi_{+}(P,\pm s) = 1 $ and $  \bar{\psi}_{+}(P,\mp s) \psi_{+}(P,\pm s) = 0  $. These lead us to write
\begin{align}
     N = N_{-} =N_{+} = \sqrt{\dfrac{\varepsilon  +mc^2}{2mc^2}} .
\end{align}
It implies in writing $\psi^{\dagger}_{+}(P,\pm s) \psi_{+}(P,\pm s)  = \varepsilon /mc^2 $  and $\psi^{\dagger}_{+}(P,\mp s) \psi(P,\pm s)  = 0$.

In the context of the negative energy solution, where $\mathscr{E} = -\varepsilon$ for both spin states, we find for spin up
\begin{align}
\psi_{-}(P,+s) &= v(P,+s) e^{iP_{\mu}x^{\mu}/\hbar}  \nonumber \\
&= N'_{+} \begin{pmatrix}  \dfrac{-c\left(P_3 +im_5 c\right)}{\varepsilon +mc^2}  \\[0.3cm]   \dfrac{-c\left(P_1+iP_2 \right)  }{\varepsilon +mc^2} \\[0.3cm] 1 \\ 0  \end{pmatrix}  e^{iP_{\mu}x^{\mu}/\hbar}
\end{align}
and for spin down
\begin{align}
\psi_{-}(P,-s) &= v(P,-s) e^{iP_{\mu}x^{\mu}/\hbar}   \nonumber \\
& = N'_{-} \begin{pmatrix}  \dfrac{-c\left(P_1-iP_2 \right)}{\varepsilon  +mc^2}  \\[0.3cm]   \dfrac{-c\left(-P_3+im_5 c \right)}{\varepsilon  +mc^2} \\[0.3cm] 0 \\ 1  \end{pmatrix} e^{iP_{\mu}x^{\mu}/\hbar} .
\end{align}
Through the requirements of $\bar{\psi}_{-}(P,\pm s) \psi_{-}(P,\pm s) = -1 $ and $ \bar{\psi}_{-}(P,\mp s) \psi_{-}(P,\pm s) = 0 $, one is able to establish that 
\begin{align}
     N = N'_{-} =N'_{+} = \sqrt{\dfrac{\varepsilon  +mc^2}{2mc^2}}  .
\end{align}
This is reflected in $\psi^{\dagger}_{-}(P,\pm s) \psi_{-}(P,\pm s)  = \varepsilon /mc^2$ and $\psi^{\dagger}_{-}(P,\mp s) \psi_{-}(P,\pm s)  = 0$.

\subsection{The Klein Paradox in presence of LSV }

\begin{figure}
    \centering
    \includegraphics{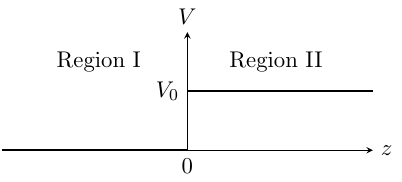}
    \caption{Potential $V = V_0\theta(z)$, in which we propose an electron traveling from the left, Region I, to the right, Region II. This potential will assist us in studying the Klein Paradox in the context of the modified Dirac equation \eqref{simplesDirac}, which we are considering in this section.}
    \label{potential}
\end{figure}

To explore the Klein paradox within the context of Eq. \eqref{simplesDirac}, it is necessary to examine the stationary wave solutions for positive energy and spin-up states, as described by Eq. \eqref{pspinup}. These solutions represent particle configurations moving along the increasing direction of the $z$ axis defined in FIG. \ref{potential}. When this is done, the particles will encounter a step potential, resulting in
\begin{align}
 V = V_0 \theta(z)=\begin{cases} 0 \,\,\, \textrm{ if } z < 0  \textrm{,}\\
                 V_0 \textrm{ if } z> 0 \textrm{.} 
    \end{cases} 
\end{align}
The graphical representation of this potential can be found in FIG. \ref{potential} as well.
Given this physical setup, one must employ a spinor $\psi_1$ to describe the free particles in Region I traveling to the Region II, which will interact with the potential $V=V_0$ in $z=0$. One also need a spinor $\psi^r_1$ to describe particles reflected by the potential in Region I. Last but not least, one should take a spinor $\psi_2$  which must describe the particles that penetrate the potential and are found in the Region II.

Henceforth, we shall proceed with the perspective we have established for effective energies or momenta, such that $\psi_1$ possesses a momentum of $\bm{P}_1 = P\hat{\bm{z}}$, $\psi_1^r$ is associated with a momentum $\bm{P}_1^r = -P\hat{\bm{z}}$, and $\psi_2$ carries a momentum of $\bm{P}_2= P'\hat{\bm{z}}$. In light of this, it is imperative to express the solutions to the modified Dirac Eq. \eqref{simplesDirac} in the form of plane waves
\begin{align}
\psi_1 =&\; A\begin{pmatrix}
 1 \\0 \\ \dfrac{-im_5c^2+cP}{\mathscr{E}+mc^2} \\ 0
\end{pmatrix} e^{iPz/\hbar}, 
\end{align}
\begin{align}
 \psi_1^{r} =&\; B\begin{pmatrix}
 1 \\0 \\ \dfrac{-im_5c^2-cP}{\mathscr{E}+mc^2} \\ 0
\end{pmatrix} e^{-i Pz/\hbar}  \nonumber \\
&\;+ D\begin{pmatrix}
 0 \\1 \\ 0\\ \dfrac{-im_5c^2+cP}{\mathscr{E}+mc^2} 
\end{pmatrix}e^{-iPz/\hbar}  ,
\end{align}
with  dispersions relations which imply in  write $Pc = \sqrt{\mathscr{E}^2 -m^2c^4 - m_5^2c^4} $, and
\begin{align}
\psi_2 =&\; C\begin{pmatrix} 
 1 \\0  \\ \dfrac{-im_5c^2+cP'}{(\mathscr{E}-V_0)+mc^2} \\0
\end{pmatrix} e^{iP'z/\hbar}  \nonumber \\
&\; + F\begin{pmatrix}
 0 \\1 \\ 0\\ \dfrac{-im_5c^2-cP'}{(\mathscr{E}-V_0)+mc^2} 
\end{pmatrix}e^{iP'z/\hbar}  ,
\end{align}
with  $P'c= \sqrt{(\mathscr{E}-V_0)^2 -m^2c^4 - m_5^2c^4} $.
 The spinor $\psi_r$ is expressed as a sum of contributions from both spin up and spin down states to accommodate the possibility of a spin-flip resulting from reflection at $z=0$. Similarly, $\psi_2$ may also undergo a spin-flip due to its interaction with the potential in Region II.

It is worth noting that when $|\mathscr{E}-V_0| < c\sqrt{m^2c^2 + m_5^2 c^2}$, the momentum in Region II becomes imaginary, denoted as $P' = i|P'|$, and the solution exhibits a decreasing exponential behavior, indicating a damped penetration into the potential region. Conversely, in situations where $V_0 > \mathscr{E} + c\sqrt{m^2c^2 + m_5^2 c^2}$, the potential effectively confines the particle, resulting in a real momentum, and leading to oscillatory solutions within Region II.

Now one should  require that in $z=0$ the solutions, inside and outside the potential, ought to be equals, \begin{align}
\psi_1(z=0)+\psi_1^{r}(z=0)=\psi_2(z=0)\textrm{.}
\end{align}
This condition gives rise to the following outcome
\begin{align}
&D=F, \label{2} \\
&A+B= C, \label{1}\\
& \dfrac{(P-im_5 c)D}{\mathscr{E}+mc^2} = - \dfrac{(P'-im_5c)F}{\mathscr{E}-V_0+mc^2}, \label{3} \\
&\dfrac{P(A-B) -im_5c(A+B) }{\mathscr{E}+mc^2} =  \dfrac{(P'-im_5c)C}{\mathscr{E}-V_0+mc^2} . \label{4}
\end{align}
The relations \eqref{2} and \eqref{3} cannot be satisfied at same time, suggesting $D=F=0$ and eliminating the spin-flip possibility. Replacing the expressions for $Pc$ and $P'c$ in equality \eqref{4} and after using the constraint  that comes from \eqref{1}, one can write
\begin{align}
A-B=  (\beta- i\alpha)C, \label{5}
\end{align}
with 
\begin{align}
&\alpha  = \dfrac{m_5c^2 V_0}{ \left( V_0 -\mathscr{E}-mc^2\right)\left(\mathscr{E}^2-m^2c^2-m_5^2 c^2 \right)^{1/2}} , \\
&\beta =\dfrac{\left(\mathscr{E}+mc^2 \right) \left[\left(\mathscr{E}-V_0\right)^2-m^2c^2-m_5^2 c^2 \right]^{1/2}}{ \left(\mathscr{E}- V_0 +mc^2\right)\left(\mathscr{E}^2-m^2c^2-m_5^2 c^2 \right)^{1/2}} .
\end{align}
Adding \eqref{1} with \eqref{5} and after making \eqref{1} minus \eqref{5} one gets, respectively,
\begin{align}
A = (1+\beta-i\alpha)\dfrac{C}{2}, \;\;\;\;\;
B =(1-\beta+i\alpha)\dfrac{C}{2}.
\end{align}
From this,  one can conclude that
\begin{align}
\dfrac{B}{A} = \dfrac{1-\beta+i\alpha}{1+\beta-i\alpha} , \;\;\;\;\;
\dfrac{C}{A} = \dfrac{2}{1+\beta-i\alpha} .
\end{align}

Finally, to fully specify the transmission and reflection coefficients, one ought to
consider the current term in the $\bm{\hat{z}}$ direction, 
\begin{align}
 \bm{j} &= c\psi^{\dagger}(z)\gamma^0 \gamma^3 \psi(z) \bm{\hat{z}} = c  \psi^{\dagger}(z)\begin{pmatrix}
           0 & \sigma_3 \\  \sigma_3 & 0
 \end{pmatrix}   \psi(z) \bm{\hat{z}},
\end{align}
in the Dirac representation for gamma matrices with $\sigma_3$ being the third Pauli matrix.   Consequently, one can show that
\begin{align}
    \bm{j}_1   &= AA^{*} \left( \dfrac{2c^2P}{\mathscr{E}+mc^2} \right) \bm{\hat{z}}  ,\\
     \bm{j}_1^r &= -BB^{*}  \left( \dfrac{2c^2P}{\mathscr{E}+mc^2} \right) \bm{\hat{z}}    , \\
     \bm{j}_2  &= CC^{*}  \left( \dfrac{2c^2P'}{(\mathscr{E}-V_0)+mc^2} \right)  \bm{\hat{z}}  .
\end{align}
Hence, the reflection coefficient shall be
\begin{align}
R &= \dfrac{| \bm{j}_1^r|}{|\bm{j}_1 |}  = \dfrac{|-BB^{*} |}{|AA^{*} |}=  \dfrac{1-2\beta + \beta^2 + \alpha^2}{(1+\beta)^2 + \alpha^2}, 
\end{align}
and the transmission coefficient is 
\begin{align}
T&= \dfrac{| \bm{j}_2|}{| \bm{j}_1 |} =\dfrac{|CC^{*} |}{|AA^{*} |} \!\bigg\vert \! \dfrac{P'}{P} \!\left(\! \dfrac{\mathscr{E}\!+\!mc^2}{\mathscr{E}\!-\!V_0\! +\!mc^2} \! \right) \!\! \bigg \vert  \nonumber \\
& =\dfrac{4\beta}{(1+\beta)^2 + \alpha^2} .
\end{align}
One can immediately verify that $R+T=1$. 

By looking at the explicit solutions worked out to study the Klein paradox, we may note that the $m_5$ parameter has a nontrivial consequence on the the transmission and reflection coefficients. Actually, by setting $m_5=0$ would considerably change the expressions  for the  transmission and reflection coefficients. We would like finally to stress that though it would appear
that $m_5$ could be absorbed into a redefinition of an effective mass, through the combination $(m^2 + m_5 ^2) c^2$, this is not the case whenever we consider the $\alpha$-coefficient,  which is explicitly present in the expressions for the reflection and transmission coefficients, as mentioned above.

As we demand the case for high barrier, $V_0> \mathscr{E}+c\sqrt{m^2c^2+m_5^2 c^2} $, we have $\beta <0$. As a result,  the transmission function in this situation is also less than zero, $T<0$.
These results suggest to us that $|\bm{j}_1^r| > |\bm{j}_1|$. The physical explanation for this phenomenon is the presence of particles moving from Region II to Region I. However, from the very beginning, we have argued that there are no particles in Region II. Therefore, it becomes apparent that the Klein's paradox is occurring within the scenario of the modified Dirac Eq. \eqref{simplesDirac}.

One commonly suggested solution to this problem is its association with pair production when the inequality $V_0> \mathscr{E} + c\sqrt{m^2c^2+m_5^2 c^2} > 2 c\sqrt{m^2c^2+m_5^2 c^2}$ holds true. This inequality implies that the potential is potent enough to create particle-antiparticle pairs from the vacuum. The difficulty here is in trying to comprehend a multiparticle phenomenon using a simple single-particle wave function. However, quantum field theory offers a viable approach for a proper treatment. 

\subsection{Implications of LSV on the \textit{Zitterbewegung}}

The term \textit{Zitterbewegung} originates from the German  and can be loosely translated into English as referring to a trembling or agitated movement. In our context, \textit{Zitterbewegung} is consistently associated with the hypothesis of a rapid intrinsic oscillation of elementary particles, whose dynamics are governed by the quantum-relativistic wave equation. This concept first emerged in 1928 with  G. Breit \cite{breit28} and after in 1930 as a direct consequence of the free solutions of the Dirac equation for wave packets, in a work by E. Schrödinger \cite{schrodinger30}. Schrödinger observed an interference phenomenon between positive and negative energy states, resulting in an electron appearing to oscillate around its classical trajectory at a natural frequency of $\omega = 2mc^2/\hbar$. Details of the original Schrödinger derivations can be found in the work by Barut and Bracken \cite{barut81}.  {Another classical reference showing the \textit{Zitterbewegung} effect for Dirac fermions is the work of Ref. \cite{foldy50}.} In our days, experimental investigations of the \textit{Zitterbewegung} phenomenology has  demonstrated remarkable progress \cite{naturezitter} in the interface with CMP scenarios. We highlight situations like graphene \cite{graphene0, graphene1, graphene2}, semiconductors \cite{semicond}, topological phase transitions \cite{toplogical1} and Bose-Einstein condensation \cite{condensation1, condensation2}. Also, in atomic physics \cite{atomic1, atomic2} and astrophysics \cite{astro1}, \textit{Zitterbewegung} has been contemplated with noticeable interest. 
 In this Section, we are going to re-assess the \textit{Zitterbewegung} in the framework of Eq. \eqref{simplesDirac}.

In order to reveal the  \textit{Zitterbewegung} frequencies in the context of Eq. \eqref{simplesDirac}, we define a wave packet 
\begin{align}
\Psi(\vec{x},t) =&\;  \int \dfrac{d^3\bm{P}}{(2\pi\hbar)^{3/2}} \sqrt{\dfrac{mc^2}{\varepsilon(\bm{P})}}  \nonumber \\
&\; \times \sum_{\pm  s} \Big[b(P,s)u(P,s)e^{-iP^{\mu}x_{\mu}/\hbar}   \nonumber \\
&\; +  d^{*}(P,s)v(P,s)e^{+iP^{\mu}x_{\mu}/\hbar}\Big] ,\label{pac1}
\end{align}
where  $b(P,s)$ and $d^*(P,s)$ are amplitudes used to construct the wave packet, $s$ carries the spin content, up or down, and the rest of the notation is self explanatory. The normalization condition implies that $ \int d^3 \bm{x} \, \Psi^{\dagger}(\bm{x},t) \Psi(\bm{x},t) = 1 $ and imposing \eqref{pac1} to satisfy it, one can write 
\begin{align}
\int d^3 \bm{x} & \Psi^{\dagger}(\bm{x},t) \Psi(\bm{x},t)  =  \nonumber \\
 &= \int  d^3\bm{P} \sum_{\pm s} \Big[ |b(P,s)|^2+ |d(P,s)|^2 \Big] = 1 , \label{result}
\end{align}
where we have used the following normalization for the Dirac delta function
\begin{align}
 \int \dfrac{ d^3 \bm{x} }{(2\pi\hbar)^3} e^{\pm i(\bm{P}\pm\bm{P}')\cdot \bm{x}/\hbar} =\delta^{(3)} \left( \bm{P}\pm\bm{P}' \right) 
\end{align}
and the fact that  $ P_0 (\bm{P}) = P'_0(\bm{P})$ to obtain the result \eqref{result}.
Furthermore, we define the  current term
\begin{align}
J^\mu &= c \int d^3 \bm{x}  \bar{\Psi} (\bm{x},t) \gamma^\mu \Psi (\bm{x},t) ,
\end{align}
and take the $i-th$ component of the Gordon decomposition,
\begin{align}
c&\bar{\Psi} (\bm{x} ,t) \gamma^i \Psi (\bm{x} ,t) = \nonumber \\
&= \dfrac{1}{2m} \left[\bar{\Psi} \left(\hat{P}^i \Psi \right)- \left(\hat{P}^i \bar{\Psi} \right)\Psi -2i \hat{P}_{\nu} \left(\bar{\Psi} \Sigma^{i\nu} \Psi\right) \right] ,
\end{align} 
with this we can write the $i-th$ component of the current 
    \begin{align}
J^i &= \int d^3\bm{P} \Bigg\{ \sum_{\pm  s}  \dfrac{P^ic^2}{\varepsilon(\bm{P})} \Big[|b(P,s)|^2+ |d(P,s)|^2 \Big] \nonumber \\
& \!\!\!\!\! + \! ic \!\!\!\sum_{\pm  (s,s') }  \!\! \Big[ b^{*}(-P,s')d^{*}(P,s)\bar{u}(-P,s')\Sigma^{i0} v(P,s) e^{2iP_0x_0 \! / \! \hbar} \nonumber \\
& \!\!\!\!\! -d(-P,s')b(P,s)\bar{v}(-\mathscr{P},s') \Sigma^{i0} u(P,s) e^{-2iP_0x_0 \! / \!\hbar} \Big] \bigg\} .\label{currentzitter}
\end{align}

The two terms proportional to exponential in Eq. \eqref{currentzitter}  are solutions that depict an interference between positive and negative energy states, a phenomenon referred to as \textit{Zitterbewegung}. These components oscillate with an explicit time dependence at a frequency determined by
\begin{align}
\omega_z \!=\! \dfrac{2P_0 c}{\hbar} \!=\! \dfrac{2c\sqrt{\bm{P}^2+m^2c^2+m_5^2c^2} }{\hbar} . %\! >\! \dfrac{2c\sqrt{m^2c^2+m_5^2}}{\hbar}\!.
\end{align}
This frequency must be higher than a frequency threshold defined on the rest energy scale, that is,
\begin{align} \label{wzlb}
    \omega_z > \dfrac{2c\sqrt{m^2c^2+m_5^2 c^2}}{\hbar} .
\end{align}

Though the \textit{Zitterbewegung} frequency involves the combination $(m^2  +  m_5 ^ 2) c^2$, which suggests the absorption of the $m_5$-parameter into a redefinition of the electron mass, we call
into question our inspection of the Klein paradox, where $m_5$ plays a role independently from the electron mass, $m$. Adopting then this viewpoint, we argue that we can proceed further by considering $m_5$ as a genuine independent parameter and we
can then constrain it by making use of the expression \eqref{wzlb} for the threshold frequency. It is our re-assessment of the Klein paradox in presence of $m_5$ the argument we follow to claim that this parameter should not always be removed upon a field redefinition.

In Ref. \cite{graphene0}  the authors  explored the phenomenon of \textit{Zitterbewegung} in a graphene superlattice comprising massless Dirac fermions with highly anisotropic group velocities. This characteristic arises from subjecting graphene to one-dimensional periodic potentials. By adjusting the parameters of the periodic potential, was demonstrated that the frequency of \textit{Zitterbewegung} oscillations can reach magnitudes on the order of $10^{12}$ Hz, while the amplitude may extend to hundreds of angstroms, and their decay rate can significantly decrease. The necessary parameters for the graphene superlattice can feasibly be achieved under current experimental setups, presenting an excellent opportunity for experimental investigation of the \textit{Zitterbewegung} effect. Then, based on the aforementioned picture, we can estimate an upper bound on $m_5$, which is not valid for a spacetime vacuum but rather for Condensed Matter scenarios, via Eq. \eqref{wzlb}  with {$m=0$} and considering an $\omega_z \sim 10^{12}$ Hz. Therefore,
\begin{align}
    m_5 < \dfrac{\hbar \omega_z}{2c^2} = 1.1 \times 10^{-30} \; \textrm{Kg} \cdot \textrm{m} / \textrm{s} .
\end{align}
Or
\begin{align}
    m_5 < 2,06 \times 10^{-3} \; e\textrm{V}/c^2 .
\end{align}
This upper limit is compatible with the energy and momentum scales of non-relativistic scenarios like the ones we find in Condensed Matter systems.

One can also investigate \textit{Zitterbewegung} within this framework  by analysing the effective Hamiltonian associated with the modified Dirac Eq. \eqref{simplesDirac}, see Eq. \eqref{Hamiltonian} for $b_\mu =0$,
\begin{align}
H = c\bm{\alpha} \cdot \bm{P} + \beta mc^2+ \kappa  m_5 c^2 ,
\end{align} 
where $\alpha^i = \gamma^0\gamma^i$, $\beta    = \gamma^0$ and $\kappa    =i\gamma^0\gamma_5$.

In the Heisenberg picture the temporal evolution of position is given by 
\begin{align}
\dfrac{d\bm{x}}{dt} = \dfrac{1}{i\hbar} [\bm{x},H] = c \bm{\alpha} \label{xevo}
\end{align} 
for a constant background. Now, the time evolution of $\bm{\alpha}$ is given by
\begin{align}
\dfrac{d\bm{\alpha}}{dt} = \dfrac{1}{i\hbar} [\bm{\alpha},H]  &=\dfrac{i}{\hbar} \{ H, \bm{\alpha} \} - \dfrac{2i}{\hbar} \bm{\alpha} H \nonumber \\
 &= \dfrac{2i}{\hbar}  \left( \bm{P} -\bm{\alpha} H \right), \label{inte}
\end{align}
where $\{ \beta , \bm{\alpha} \} = \{\kappa, \bm{\alpha} \} = 0$. Integrating the result \eqref{inte} with respect to time, 
\begin{align}
  \int_{\bm{\alpha}(0)}^{\bm{\alpha}(t)} \dfrac{d\bm{\alpha}}{\bm{\alpha} - \dfrac{c}{H} \bm{P}} = -\dfrac{2iH}{\hbar}\int_{t'=0}^{t'=t} dt' ,
\end{align}
one attains
\begin{align}
  \bm{\alpha} (t)= \left[\bm{\alpha} (0)- \dfrac{c}{H} \bm{P}\right] e^{-2iH/\hbar} +  \dfrac{c}{H} \bm{P}  \textrm{.} \label{Z2}
\end{align}
Taking the Eq. \eqref{Z2} to the Eq. \eqref{xevo}, one arrives at the following result
\begin{align}
\bm{x}(t)\!=\!\bm{x}(0) \!+\!  \dfrac{c}{H} \bm{P} t\! +\!\dfrac{i\hbar c}{2H}\! \left(\!\bm{\alpha} (0)- \dfrac{c}{H} \bm{P}\! \right)\!\!\left(\! e^{-2iH/\hbar} \!-\! 1\!\right) .
\end{align}

Hence, we find that the motion of a fermionic particle is described by an initial position term followed by a velocity term multiplied by time. This velocity term adopts the shape of the group velocity and is accompanied by an oscillatory component responsible for \textit{Zitterbewegung}. This oscillatory behavior disrupts the typical kinematic structure of the equation of motion. In essence, we observe a fermion oscillating between positive and negative energy states as it traces its classical-type trajectory.

\section{Proposing an asymmetry between space and time for Dirac fermions} \label{sec2}

The efforts of this Section are motivated by a theoretical exploration inspired by H. Isobe and N. Nagaosa's study of quantum critical phenomena in the phase transition between trivial and topological insulators in $(3+1)$D. Their study involves a Dirac fermion coupled to the electromagnetic field \cite{isobe2012theory}. In the model they considered, LSV arises due to a constant dimensionless parameter that, from the beginning, only affects the velocity term of the Dirac Hamiltonian. Our proposal involves constructing an $U(1)$ gauge theory based on an extended version of the Isobe-Nagaosa model, wherein the LSV parameter becomes space-time dependent.

As far as the asymmetry between the time and space components of the kinetic Dirac term
is concerned, we adopt the viewpoint taken by Isobe and Nagoasa in their paper \cite{isobe2012theory}. They
actually demonstrate that the asymmetry, in our case expressed by the parameter $\xi$, is
physically relevant. Based on that, we argue that it could be of interest to inspect the
possibility that $\xi$ be in a general case space-time dependent. This means that the asymmetry
described by $\xi$ could be localized. Our true motivation to introduce $\xi$ is then justified by
the physically consequent results worked out by Isobe and Nagaosa.

To begin, we put forward a Lagrangian in a manner akin to the aforementioned work. However, we now allow for the parameter responsible for LSV, denoted here as $\xi$, to possess a nontrivial space-time dependence. In this context, omitting the surface term, we express the Dirac Lagrangian in the following manner
\begin{align}
\mathcal{L} =&\; \dfrac{i}{2} \dfrac{\hbar}{c} \bar{\psi} \gamma^0 \left(\partial_t \psi \right) -\dfrac{i}{2} \dfrac{\hbar}{c} \left(\partial_t \bar{\psi} \right) \gamma^0 \psi \nonumber\\ 
&\; +\dfrac{i}{2} \xi \hbar \bar{\psi} \gamma^i \left(\partial_i \psi \right) - \dfrac{i}{2} \xi\hbar \left(\partial_i \bar{\psi} \right) \gamma^i \psi -mc \bar{\psi}\psi  \nonumber \\
            =&\;  \bar{\psi}\left(i \hbar\gamma^0 \partial_0+i \hbar\xi \gamma^i\partial_i -mc \right) \psi +\dfrac{i\hbar}{2} \left( \partial_i \xi \right) \bar{\psi} \gamma^i \psi  ,
\end{align} 
which despite rescuing the Isobe-Nagaosa model when $\xi$ is again a constant,  generates an action that is not real.

To address this inconsistency, we assume that in situations where $\xi$ is not constant, this parameter takes values in the complex numbers. In this vein, we write
\begin{align}
\mathcal{L} =&\; \dfrac{i}{2} \dfrac{\hbar}{c} \bar{\psi} \gamma^0 \left(\partial_t \psi \right) -\dfrac{i}{2} \dfrac{\hbar}{c} \left(\partial_t \bar{\psi} \right) \gamma^0 \psi \nonumber \\
&\; + \dfrac{i}{2} \xi \hbar \bar{\psi} \gamma^i \left(\partial_i \psi \right) - \dfrac{i}{2} \xi^*\hbar \left(\partial_i \bar{\psi} \right) \gamma^i \psi    -mc \bar{\psi}\psi\nonumber \\
            =&\;  \bar{\psi}\!\left[\!i\hbar\gamma^0 \partial_0\! + \!\dfrac{i\hbar}{2}\! \left(\xi +\xi^* \right)\!\gamma^i \partial_i \!-\!mc  \right]\!\! \psi \!+\!\dfrac{i\hbar}{2} \!\left( \partial_i  \xi^* \right)  \!\bar{\psi} \gamma^i \psi .
\end{align}
It is important to emphasize that the parameter $\xi$ cannot be reabsorbed at the level of the action because we consider it
space-time dependent and then it cannot be eliminated by field redefinition. 
To ensure the condition of reality is met, we must stipulate that
\begin{align}
    \xi\!+\!\xi^* \!=\! \textrm{Re}(\xi)\! + \!i\textrm{Im}(\xi) \!+\! \textrm{Re}(\xi)\! -\! i\textrm{Im}(\xi) \!=\! 2\textrm{Re}(\xi) \!\in\! \mathbb{R} 
\end{align}
and 
\begin{align}
       \partial_i \xi^* = \partial_i \left(\textrm{Re}(\xi) - i\textrm{Im}(\xi) \right) \! \in \!\mathbb{C} .
\end{align}
For these two conditions to be fulfilled concurrently, it's necessary for $\textrm{Re}(\xi)$ to stay constant, with $\textrm{Im}(\xi)$ accounting for the complete space-time dependency of the parameter $\xi$. In these terms, defining $\textrm{Re}(\xi)\equiv R$ and $\textrm{Im}(\xi) \equiv I(x)$, one can write the Lagrangian that results in a real action as follows
\begin{align} \label{lagra1}
    \mathcal{L}= \bar{\psi}\left[i\hbar\gamma^0 \partial_0 + i\hbar R \gamma^i \partial_i-mc  \right] \psi +\dfrac{\hbar}{2} \left(\partial_i I \right)  \!\bar{\psi} \gamma^i \psi .
\end{align}
Defining $ \tilde{\partial}_\mu \equiv (c^{-1}\partial_t, R \bm{\nabla}) $, the modified Dirac equation associated with the Lagrangian \eqref{lagra1} can be expressed as
\begin{align} \label{moddiracR}
    \left[ i\hbar \gamma^\mu \tilde{\partial}_\mu -mc +\dfrac{\hbar}{2}(\partial_i I)\gamma^i \right] \psi= 0. 
\end{align}

{
To implement the interaction between charged fermions, as described by Eq. \eqref{moddiracR}, with photons, we follow, as usually, the minimal coupling prescription. This means that we have to gauge-covariantize the spacetime derivatives according to what follows:
\begin{align}
    \partial_0 &\to \partial_0 +i\dfrac{qe}{\hbar} A_0 , \\
    \partial_i &\to \partial_i -i \dfrac{qe}{\hbar} A_i,
\end{align}
where $qe$ is a multiple $q$ of the fundamental electrical charge, $e$, whereas $A_0$ and $A_i$ are the scalar and $i$ component vector potentials. With that done, the Lagrangian \eqref{lagra1} takes the form 
\begin{align}
    \mathcal{L} = \bar{\psi}\left[i\hbar \gamma^\mu \tilde{\partial}_\mu +\dfrac{\hbar}{2}(\bm{\nabla} I)\cdot \bm{\gamma} -mc\right]\psi -qe A_\mu \bar{\psi}\tilde{\gamma
}^\mu \psi, 
\end{align}
defining  $\tilde{\gamma
}^\mu \equiv (\gamma^0,R\bm \gamma)$ we can define a modified current term as 
\begin{align}
   \tilde{J}^\mu \equiv (J^0, R \bm J) .
\end{align}
}

\subsection{Expanding the background $I(x)$}
\label{sec3A}

{
The anisotropy $\partial_i I(x)$ will be known once the $I$-function is given. It turns out that, in the case this function is non-linear in the spacetime coordinates, one would introduce an explicit dependence on $t$ and$x^i$ in the Dirac equation \eqref{moddiracR}; this would play the role of an external potential the electron is subject to, explicitly breaking spacetime translational symmetry and leading to energy and linear momentum non-conservation of the charged fermion-photon system.} 

{
To avoid explicit powers of $x^i$ in the equation \eqref{moddiracR}, we then take $I(x)$ with a linear space-time dependence as follows:
\begin{align} \label{dependence}
    I(x) = \alpha + w_\mu x^\mu ,
\end{align}
with $\alpha$ and $w_\mu$ respectively being a dimensionless constant scalar and a four-vector with inverse length dimension.
}

{
 With that ansatz, one can write $w_\mu x^\mu = cw_0 t -\bm{w}\cdot \bm{x} $ and as a result  
\begin{align}
    \partial_i I(x) = -\partial_i w_j x_j =- w_j \delta_{ij} = -w_i.
\end{align}
With this statement, the modified Dirac Eq. \eqref{moddiracR} in the momentum space takes the form 
\begin{align} \label{momspace1}
     \left(\tilde{p}_\mu\gamma^\mu -mc -\dfrac{\hbar}{2} w_i \gamma^i \right) \psi(p) = 0 ,
\end{align}
with $\tilde{p}_\mu \equiv (E/c, - R \bm p)$. Therefore, the term $w_i\gamma^i$ can be mapped to the spatial component of  $a_\mu \gamma^\mu$ in the SME action given by Eq. \eqref{01}. Then, as a consequence of the discussion on removing $a_\mu$ through a field reparametrization in Sect. \ref{sec1}, we can conclude that $w_i$ has no physical consequences if the $I$-funtion is taken to be linear in the space coordinates.
}

{
The effect of the reparametrization on the components of the momentum does not change the form of the fermion propagator, but is affected only the $R$-parameter, which, as already mentioned, is constant:
\begin{align}
    iS(p) = \dfrac{i(\tilde{p}\gamma^\mu +mc)}{\tilde{p}^2 -m^2c^2} .
\end{align}
The same applies to the dispersion relation 
\begin{align} \label{dr22}
    E = \pm c \sqrt{ R^2 \bm p^2 +m^2c^2} 
\end{align}
 group velocity in the form 
\begin{align}
    \bm v = \dfrac{c^2R^2 \bm p}{E}  .
\end{align}
}

{
In the next Subsection, after we work out and get the wavelength change of a photon in a Compton scattering process by an electron, the physical consequence of the constant $R$-parameter shall be properly clarified in connection with the breaking of the linearity on the wavelengths of both the incoming and scattered photons.}

{
To our sense, in a different scenario, as it can be found in work of Ref. \cite{isobe2012theory}, which addresses to a Condensed-Matter system, a physical consequence of the constant parameter $R$ can also be justified.
}

\subsection{The kinematics of the Compton effect}

\begin{figure}[h]
    \centering
    \includegraphics{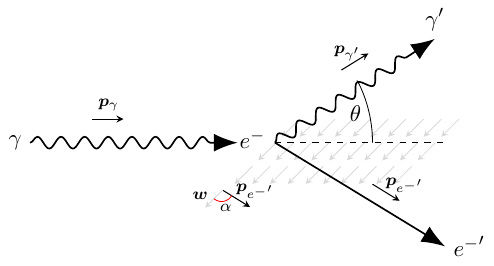}
    \caption{Compton scattering of a photon, $\gamma$, interacting with a stationary electron, $e^-$, producing a scattered electron, ${e^-}'$, and a photon, $\gamma'$. {
    This process could, in principle, happen in an anisotropic space due to LSV constant vector $\bm w$. But as shown in Sect. \ref{sec3A}, this type of Lorentz symmetry violation would not have observable effects on scattering; only $R$ would.
    } 
    }
    \label{compton}
\end{figure}

{
In this Section, let us consider the Compton scattering kinematics as illustrated in FIG. \ref{compton}, where a photon, $\gamma$, interacts with a stationary electron, $e^-$; as a result of this interaction, we have a photon $\gamma'$ and a scattered electron $e^{'},$ with the direction of propagation of $\gamma'$ making an angle $\theta$ with the incoming direction of $\gamma$, see FIG. \ref{compton}. In this regard, the rest energy of the incoming electron  is kept unchanged and the dispersion relation for the outgoing electron is modified by Eq.  \eqref{dr22}. 
}

{We would like to clarify that LSV effects are present in both the fermion and photon propagations. We adopt the point of view of considering modifications just on the electron dispersion relations. We justify  this choice because we are interested in keeping track of LSV exclusively in the fermion sector. If the modified photon dispersion relation is simultaneously introduced, LSV parameters from both the photon and fermion sectors interfere and we cannot isolate the pure effects of the parameters brought about by the fermion.
}

{
Initially,  we assume that the incoming energies of the photon and electron are respectively given by $E_{\gamma} = \omega\hbar$ and $E_{{e^-}} =  m_e c^2$.
The scattering occurs and final energy of the system takes the form $E_{\gamma}' = \omega'\hbar$ and  $ E_{{e^-}'}  =c \sqrt{  R^2\bm{p}_{{e^-}'} ^2  +m_e^2c^2} $.
The conservation of energy and linear momentum entails that
\begin{align}
   &E_{\gamma} + E_{e^{-}} =  E_{\gamma'} + E_{{e^-}'} \label{energy1}\\
 &\bm{p}_{\gamma} = \bm{p}_{\gamma'} + \bm{p}_{{e^-}'}. \label{mome1}
\end{align}
}

{Solving \eqref{energy1} for $(c| \bm{p}_{{e^-}'}|)^2$, where $|\bm{p}_{{e^-}'}|$ and $|\bm w|$ are the electron linear momentum,  one can write  
\begin{align}
    (c| \bm{p}_{{e^-}'}|)^2 =&\; \dfrac{1}{R^2}\left\{ \left[(\omega\hbar -\omega'\hbar) +m_e c^2 \right]^{2}  -m_e^2 c^4  \right\}, \label{res1}
\end{align}
}

{
One can also rewrite Eq. \eqref{mome1} as $\bm{p}_{{e^-}'} = \bm{p}_{\gamma} - \bm{p}_{\gamma'}$ and by taking the scalar product of this equation with $c^2 \bm{p}_e$, one can express
\begin{align}
    c^2 | \bm{p}_{{e^-}'}|^2 &= c^2 |\bm{p}_\gamma|^2 + c^2 |\bm{p}_{\gamma '}|^2 -2 c^2 |\bm{p}_\gamma||\bm{p}_{\gamma '}| \cos (\theta) \nonumber \\
    &= (\omega\hbar)^2+(\omega'\hbar)^2 -2(\omega\hbar)(\omega'\hbar)\cos (\theta) , \label{res2}
\end{align}
where $\theta$ is the angle between the incoming and outgoing photons. 
}

{
To finish, one ought to guarantee the equality between the right-hand sides of results \eqref{res1} and \eqref{res2}. If it so, and introducing the relations $\omega =2\pi \nu$ and $\nu= c/\lambda$ as well as the definition of Compton wavelength, $\lambda_c = h/m_ec$, and $\Delta \lambda = \lambda'-\lambda$, one obtain  a modified expression for the Compton scattering in the form 
\begin{align} \label{scatt}
  \Delta \lambda \left[1+ \dfrac{ \lambda_c}{2\lambda' \lambda} (1-R^2) \Delta \lambda \right]= R^2 \lambda_c (1- \cos \theta).
\end{align}
}

{
This result reduces to the usual expression for the Compton scattering, $\Delta \lambda =\lambda_c (1- \cos \theta)$, when $R=1$. As we can see explicitly, while in the usual Compton scattering the wavelength shift $\Delta \lambda$ depends on just the angle $\theta$, when the LSV effect of $R$ parameter comes to play this is not true anymore, with $R$ contributing to the non-linear relation between wavelengths of the incoming
and outgoing photons in \eqref{scatt}.
}

\section{Concluding Remarks } \label{sec4}

In this contribution, we have investigated the implications of a Lorentz symmetry breakdown within two distinct contexts: We investigate the fermionic Lorentz- and CPT-violating sector of QED within the SME framework \cite{Kostelecky1}. Specifically, we focus on terms of the form $M = m + \delta M_{\textrm{\tiny LSV}}  $, as outlined in Eq. \eqref{MMterm}; ii) Additionally, we examine an extended version of the Isobe-Nagaosa model \cite{isobe2012theory}, wherein the traditional hierarchy between space and time is disrupted. This disruption arises from a local Lorentz symmetry-violating scale factor that modifies the spatial part of the four-derivative within the Dirac equation.  

In the initial investigation, we re-examined the modified Dirac equation and then derived the extended fermion propagator, group velocity, and dispersion relation,  emphasizing the criteria for ensuring causality. Additionally, we derived the Gordon decomposition and highlighted that the $b_\mu$ LSV term introduces a current component linked to the electromagnetic form factor associated with the Electric Dipole Moment (EDM). Taking a step forward, we argue that the canonical energy-momentum tensor, when LSV terms are present, aligns with that of conventional Dirac theory, along with its conservation law  upon the assumption that LSV terms are held constant. Furthermore, employing an approach outlined in \cite{Freese22},  we show that LSV terms avoid the symmetrization of the canonical energy-momentum tensor and make it incapable of being conserved.

The particular case of {$b_\mu = 0 $} in Eq. \eqref{02}, in which we explored Klein's paradox which suggests  that quantum field theory offers
a viable approach for a proper treatment of creation and annihilation of particles, it turned out to be particularly interesting once it opened, via \textit{Zitterbewegung} phenomenology, a window for establishing an upper bound on the LSV $m_5$ parameter on the order of $< 10^{-3} e$V/c. This bound is anchored in the data for \textit{Zitterbewegung} frequency of the order $10^{12}$ Hz  in a graphene superlattice comprising massless Dirac fermions with highly anisotropic group velocities \cite{graphene0} and aligns with the energy and momentum scales typically observed in non-relativistic scenarios, such as those found in Condensed Matter systems.

In our proposal of constructing an $U (1)$ gauge theory based on an extended version of the Isobe-Nagaosa model with the LSV parameter carrying a space-time dependent form, firstly we are faced with the nontriviality of writing an action for the resulting Dirac equation, which was overcome by requiring this LSV parameter to take values in the complexes. Then, by considering a linear space-time dependence on this LSV parameter, we were capable to explore how this Lorentz symmetry breakdown affects the usual relativistic dispersion relation and the group velocity. Additionally, when we  consider a minimal coupling with the photon, the interaction between electromagnetic radiation and matter is affected by the LSV in consideration. In has the potential to open new avenues for exploring how this Lorentz symmetry breakdown affects the Maxwell sector and its observables. 

To end up, we briefly investigate the kinematics of the Compton effect, which plays an important role in astrophysical phenomena, if we consider that extra-galactic photons may be scattered by electrons and positrons that appear as a byproduct of the Breit-Wheeler effect produced by the scattering between gamma-rays \cite{Breitt34, Lobet17} and extra-galactic backgroung light (EBL) \cite{Dwek13, Krennrich13}. This model also motivates us to investigate, in a future work, new effects of this modified QED, such as the corrections to the $(g - 2)$ magnetic anomaly of the electron and muon, which are decisive high-precision tests for the Standard Model \cite{Crivellin22,Bluhm97c, Bluhm98c}. Another study we shall pursue consists in computing the effects of the LSV fermionic parameters on the corrections to the spin- and velocity-dependent contributions to the Coulomb potential of charged fermions. We shall report on that elsewhere in a forthcoming paper. Finally, it is also worth mentioning a recent work, whose main purpose is to understand how LSV - as initially manifest here only in the fermionic sector - influences the physics of the charged scalar partners of the fermion matter in the $N=1$- supersymmetric extension of QED \cite{JpWH23}.

\section{Acknowledgements}

J.P.S.M.  expresses his gratitude to CNPq-Brazil for granting his PhD Fellowship.

\end{document}